# Does the Poynting vector always represent electromagnetic power flow?

Changbiao Wang[*]
ShangGang Group, 70 Huntington Road, Apartment 11, New Haven, CT 06512, USA

Poynting vector as electromagnetic power flow has prevailed over one hundred years in the community. However in this paper, it is shown from Maxwell equations that the Poynting vector may not represent the electromagnetic power flow for a plane wave in a non-dispersive, lossless, non-conducting, *anisotropic* uniform medium; this important conclusion revises the conventional understanding of Poynting vector. It is also shown that this conclusion is clearly supported by Fermat's principle and special theory of relativity.

*Key words*: Poynting vector; electromagnetic power flow; plane wave in an anisotropic medium; Maxwell equations; Fermat's principle

*PACS Nos.*: 03.50.De, 41.20.-q

## 1. Introduction

An electromagnetic (EM) plane wave, although not practical, is a simplest strict solution of Maxwell equations, and it is often used to explore most fundamental physics. For example, Einstein used it to develop his special theory of relativity and derived the well-known relativistic Doppler formula [1].

In conventional EM wave theory, Poynting vector as EM power flow (the rate of flow of energy per unit area per unit time) has been thought to be a well-established basic concept [2-7]. In view of the existence of some kind of mathematical ambiguity for this concept, some scientists suggested it to be a "hypothesis", "until a clash with new experimental evidence shall call for its revision" [2]. However this conventional concept has been questioned from various perspectives [8-10]. Nevertheless, recently some scientists have proposed it to be a "postulate" for resolution of the long-lasting Abraham-Minkowski controversy [11,12]. This situation is very confusing.

In this paper, from Maxwell equations it is shown that the Poynting vector does not necessarily represent the EM power flow for a monochromatic plane wave in a non-dispersive, lossless, non-conducting, *anisotropic* uniform medium. This important conclusion revises the basic concept, which has prevailed over one hundred years. It is also shown that this conclusion is clearly supported by Fermat's principle and special theory of relativity.

## 2. Proof

As a basic concept of EM wave theory, the correctness of Poynting vector as EM power flow should withstand any tests of EM waves, especially the test of a monochromatic plane wave, because the Poynting vector consists of the EM fields satisfying Maxwell equations while a monochromatic plane wave satisfies the Maxwell equations.

In a non-dispersive, lossless, non-conducting, anisotropic uniform medium, for a monochromatic plane wave with a phase function $\Psi = (\omega t - n_d \mathbf{k} \cdot \mathbf{x})$, the Maxwell equations are simplified into

$$\omega \mathbf{B} = n_d \mathbf{k} \times \mathbf{E}, \qquad \omega \mathbf{D} = -n_d \mathbf{k} \times \mathbf{H}, \qquad (1)$$

$$(n_d \mathbf{k}) \cdot \mathbf{B} = 0, \qquad (n_d \mathbf{k}) \cdot \mathbf{D} = 0, \qquad (2)$$

where $(\mathbf{E}, \mathbf{D}, \mathbf{B}, \mathbf{H}) = (\mathbf{E}_0, \mathbf{D}_0, \mathbf{B}_0, \mathbf{H}_0) \cos \Psi$ with $\mathbf{E}_0$, $\mathbf{D}_0$, $\mathbf{B}_0$, and $\mathbf{H}_0$ being the real constant vectors, $\omega$ is the frequency, $n_d \mathbf{k}$ is the wave vector, and $n_d = |n_d \mathbf{k}|/|\omega/c|$ is the refractive index of medium, with $c$ the vacuum light speed. $\omega$ and $n_d \mathbf{k}$ are real because the medium is assumed to be lossless and non-conducting.

By making cross products of $n_d \mathbf{k} \times (\omega \mathbf{B} = n_d \mathbf{k} \times \mathbf{E})$ and $n_d \mathbf{k} \times (\omega \mathbf{D} = -n_d \mathbf{k} \times \mathbf{H})$ from Eq. (1), with vector identity $\mathbf{a} \times (\mathbf{b} \times \mathbf{c}) = (\mathbf{a} \cdot \mathbf{c}) \mathbf{b} - (\mathbf{a} \cdot \mathbf{b}) \mathbf{c}$ taken into account, we have

$$\mathbf{E} = (\hat{\mathbf{n}} \cdot \mathbf{E})\hat{\mathbf{n}} - \mathbf{v}_{ph} \times \mathbf{B} , \tag{3}$$

$$\mathbf{H} = (\hat{\mathbf{n}} \cdot \mathbf{H})\hat{\mathbf{n}} + \mathbf{v}_{ph} \times \mathbf{D} , \tag{4}$$

where $\mathbf{v}_{ph} = \hat{\mathbf{n}}(\omega/|n_d \mathbf{k}|)$ is the phase velocity, and $\hat{\mathbf{n}} = n_d \mathbf{k}/|n_d \mathbf{k}|$ is the unit wave vector.

---

[*] Email address: changbiao_wang@yahoo.com

By making inner products of $\mathbf{H} \cdot (\omega \mathbf{B} = n_d \mathbf{k} \times \mathbf{E})$ and $\mathbf{E} \cdot (\omega \mathbf{D} = -n_d \mathbf{k} \times \mathbf{H})$ from Eq. (1), with $\mathbf{H} \cdot (n_d \mathbf{k} \times \mathbf{E}) = \mathbf{E} \cdot (-n_d \mathbf{k} \times \mathbf{H})$ taken into account we have $\mathbf{E} \cdot \mathbf{D} = \mathbf{B} \cdot \mathbf{H}$. Setting $\mathbf{S} = \mathbf{E} \times \mathbf{H}$ (Poynting vector) and $W_{em} = 0.5(\mathbf{E} \cdot \mathbf{D} + \mathbf{B} \cdot \mathbf{H})$ (EM energy density), from Eqs. (3) and (4) we obtain

$$\mathbf{S} = \mathbf{S}_{power} + \mathbf{S}_{pseu} , \tag{5}$$

where

$$\mathbf{S}_{power} = \mathbf{v}_{ph}^2 (\mathbf{D} \times \mathbf{B}) = W_{em} \mathbf{v}_{ph} , \tag{6}$$

$$\mathbf{S}_{pseu} = -(\mathbf{v}_{ph} \cdot \mathbf{H})\mathbf{B} - (\mathbf{v}_{ph} \cdot \mathbf{E})\mathbf{D} , \tag{7}$$

and they are perpendicular each other ($\mathbf{S}_{power} \perp \mathbf{S}_{pseu}$).

From Eq. (6), we see that $\mathbf{S}_{power}$ carries *all* the EM energy $W_{em}$ and propagates at the phase velocity $\mathbf{v}_{ph}$. Thus $\mathbf{S}_{power}$ is the real power flow and $\mathbf{S}_{pseu}$ is the pseudo-power flow. The physical difference between $\mathbf{S}_{power}$ and $\mathbf{S}_{pseu}$ also can be seen from the divergence theorem.

From Eq. (6) we have $\nabla \cdot \mathbf{S}_{power} \neq 0$ (except for those discrete points) while the time average $<\nabla \cdot \mathbf{S}_{power}> = 0$, which means that $\mathbf{S}_{power}$ is responsible for a power flowing in and out in a differential box, and the powers going in and out are the same on time average, with no net energy left in the box.

In contrast, from Eq. (7) we have $\nabla \cdot \mathbf{S}_{pseu} \equiv 0$ holding due to $\nabla \cdot \mathbf{B} = 0$, $\nabla \cdot \mathbf{D} = 0$, $\mathbf{B} \perp n_d \mathbf{k}$, and $\mathbf{D} \perp n_d \mathbf{k}$. Mathematically, the expression $\nabla \cdot \mathbf{S}_{pseu} \equiv 0$ intself can imply that $\mathbf{S}_{pseu}$ is responsible for transporting a time-independent energy density $W_{pseu}$, namely $-\nabla \cdot \mathbf{S}_{pseu} = \partial W_{pseu}/\partial t \equiv 0$. But the total EM energy density is given by $W_{em} = 0.5(\mathbf{E} \cdot \mathbf{D} + \mathbf{B} \cdot \mathbf{H})$, and from the energy conservation, $W_{pseu} + W_{em} = W_{em} \Rightarrow W_{pseu} \equiv 0$. Thus for the plane wave, $\mathbf{S}_{pseu}$ is not responsible for any power flowing at any time for any places. That is why $\mathbf{S}_{pseu}$ is called pseudo-power flow.

In an isotropic medium (including empty space), Poynting vector is parallel to the wave vector, leading to $\mathbf{S}_{pseu} = 0$ and $\mathbf{S} = \mathbf{S}_{power}$; thus the Poynting vector always represents the power flow.

In an electro-anisotropic uniaxial medium, which is the simplest anisotropic medium, for any given propagation direction there are two kinds of waves: ordinary wave and extraordinary wave [13]. For the ordinary wave, with $\mathbf{H} \perp n_d \mathbf{k}$ and $\mathbf{E} \perp n_d \mathbf{k}$ both holding, Poynting vector is parallel to the wave vector $n_d \mathbf{k}$, leading to $\mathbf{S}_{pseu} = 0$; thus the Poynting vector represents the power flow. For the extraordinary wave except for those special cases where $n_d \mathbf{k}$ is parallel or perpendicular to the optical axis *Z*, as shown in Fig. 1, Poynting vector is not parallel to the wave vector $n_d \mathbf{k}$, leading to $\mathbf{S}_{pseu} \neq 0$; thus the Poynting vector does not represent the power flow. In short, the Poynting vector does not necessarily represent the power flow in an anisotropic medium.

Now let us examine the justification of $\mathbf{S}_{power}$ as the power flow from energy conservation. The EM energy conservation equation for a plane wave is given by

$$-\nabla \cdot \mathbf{S} = \frac{\partial W_{em}}{\partial t} . \tag{8}$$

In principle, EM field solutions can be obtained by solving Maxwell equations associated with their boundary conditions without any ambiguity. However there does be some ambiguity for the definition of power flow in terms of above Eq. (8). Traditionally, $\mathbf{S} = \mathbf{E} \times \mathbf{H}$ is defined as the power flow [2-7]. However by adding a term with a zero divergence to $\mathbf{S}$, Eq. (8) will not be affected. For example, inserting $\mathbf{S} = \mathbf{S}_{power} + \mathbf{S}_{pseu}$ into Eq. (8), with $\nabla \cdot \mathbf{S}_{pseu} \equiv 0$ taken into account we have the same-form conservation equation

$$-\nabla \cdot \mathbf{S}_{power} = \frac{\partial W_{em}}{\partial t} . \tag{9}$$

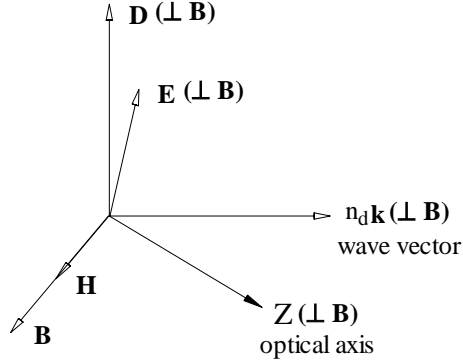

**Fig. 1.** General space relations between EM fields **E**, **D**, **B**, **H**, wave vector $n_d\mathbf{k}$, and optical axis $Z$ for extraordinary wave in an electro-anisotropic uniaxial medium [13]. $\mathbf{B} \perp n_d\mathbf{k}$ and $\mathbf{B} \perp \mathbf{E}$ from $\omega\mathbf{B} = n_d\mathbf{k} \times \mathbf{E}$. $\mathbf{H} // \mathbf{B}$ from $\mathbf{B} = \mu\mathbf{H}$, with $\mu$ the constant scalar permeability. $\mathbf{D} \perp n_d\mathbf{k}$ and $\mathbf{D} \perp \mathbf{H}$ from $\omega\mathbf{D} = -n_d\mathbf{k} \times \mathbf{H}$. $\mathbf{D}$, $\dot{\mathbf{E}}$, $n_d\mathbf{k}$, and the optical axis $Z$ lie on the same plane. $(\mathbf{D} \times \mathbf{B})//(n_d\mathbf{k})$ holds, and Poynting vector $\mathbf{S} = \mathbf{E} \times \mathbf{H}$ is not parallel to the wave vector $n_d\mathbf{k}$ except for $n_d\mathbf{k}$ // or $\perp$ the optical axis $Z$.[1]

Thus we can re-define $\mathbf{S}_{power}$ as the power flow. For an isotropic medium, $\mathbf{S}_{pseu} = 0$ and $\mathbf{S} = \mathbf{S}_{power}$, and this re-definition has no effect.

From above analysis we can see that, from the viewpoint of EM energy conservation, $\mathbf{S}$ and $\mathbf{S}_{power}$ have the equal right to be the power flow. However $\mathbf{S}$, as being EM power flow, may contain a unphysical pseudo-power flow, while $\mathbf{S}_{power}$ does not. From this perspective, it is justifiable to take $\mathbf{S}_{power}$ as the correct power flow in an anisotropic medium (crystal).

Fermat's principle is an additional physical condition imposed on the direction of EM energy transport. The medium, which supports a plane wave, is uniform ($\partial n_d/\partial \mathbf{x} = 0$), but it can be isotropic or anisotropic. According to the Fermat's principle, the optical length of an actual ray from one equi-phase plane to the next is the one to make $\int n_d ds = n_d \int ds$ the minimum [3, p.115]. When $\int ds$ is equal to the distance between the equi-phase planes, $\int n_d ds$ reaches the minimum. Thus the actual ray or the direction of energy transport must be parallel to the wave vector. $\mathbf{S}_{power}$ is parallel to the wave vector while $\mathbf{S}_{pseu}$ is perpendicular to the wave vector. If $\mathbf{S}_{power}$ is defined as the power flow, Fermat's principle is automatically satisfied. Thus $\mathbf{S}_{power}$ as power flow is also required by Fermat's principle.

It might be interesting to point out that, if $\mathbf{S}_{pseu} \neq 0$ can be of EM power flow, a "superluminal power flow" could be constructed, given by $\mathbf{S}_{>c} = \mathbf{S} + a\mathbf{S}_{pseu}$, where $a$ is an *arbitrary* constant, with $\mathbf{S}_{>c} = \mathbf{S}$ for $a = 0$ and $\mathbf{S}_{>c} = \mathbf{S}_{power}$ for $a = -1$. Obviously, $\mathbf{S}_{>c}$ satisfies energy conservation Eq. (8) due to $\nabla \cdot \mathbf{S} \equiv \nabla \cdot \mathbf{S}_{>c}$. Since $\mathbf{S}_{pseu} \perp \mathbf{S}_{power}$ holds, we have

$$\frac{|\mathbf{S}_{>c}|}{W_{em}} = \frac{1}{W_{em}}\sqrt{\mathbf{S}_{power}^2 + (a+1)^2 \mathbf{S}_{pseu}^2} . \qquad (10)$$

From this we have $|\mathbf{S}_{>c}/W_{em}| > c$ holding for $\mathbf{S}_{pseu} \neq 0$ by a proper choice of *a*-value, which, clearly, is not consistent with the special theory of relativity [1].

## 3. Conclusions

In summary, we have shown that the Poynting vector may not represent the EM power flow for a plane wave in a non-dispersive, lossless, non-conducting, *anisotropic* uniform medium, which means that the conventional understanding of Poynting vector [2-7] should be revised. However it should be emphasized that the Poynting-vector surface integral $\oiint \mathbf{S} \cdot d\mathbf{A} = \oiint \mathbf{S}_{power} \cdot d\mathbf{A}$ is always equal to the total EM power, no matter whether in an isotropic or anisotropic dielectric medium.

---

[1] Note: When the wave vector $n_d\mathbf{k}$ is parallel or perpendicular to the optical axis $Z$, we have $\mathbf{E}//\mathbf{D}$ holding, resulting in $\mathbf{S} = \mathbf{E} \times \mathbf{H}$ parallel to $n_d\mathbf{k}$, which is shown as follows. Since $\mathbf{D} = \bar{\bar{\varepsilon}} \cdot \mathbf{E}$ with $\bar{\bar{\varepsilon}} = diag(\varepsilon, \varepsilon, \varepsilon_z)$ the constant permittivity diagonal tensor [13], we have $\mathbf{D} = \varepsilon \mathbf{E}_x + \varepsilon \mathbf{E}_y + \varepsilon_z \mathbf{E}_z$. From $\nabla \cdot \mathbf{D} = 0$, we have $\mathbf{D} \perp n_d\mathbf{k}$ holding. (*a*) If $n_d\mathbf{k}//Z$ holds, we have $\mathbf{D} \perp n_d\mathbf{k} \Rightarrow \mathbf{D} \perp Z \Rightarrow \mathbf{D}_z = 0 \Rightarrow \mathbf{E}_z = 0$ $\Rightarrow \mathbf{D} = \varepsilon\mathbf{E}_x + \varepsilon\mathbf{E}_y = \varepsilon\mathbf{E}$; namely $\mathbf{E}//\mathbf{D}$ holds for $n_d\mathbf{k}//Z$. (*b*) If $n_d\mathbf{k} \perp Z$ holds, we have $\mathbf{D} \perp n_d\mathbf{k} \Rightarrow \mathbf{D}//Z$ because $\mathbf{D}$, $n_d\mathbf{k}$, and $Z$ lie on the same plane for the extraordinary wave required by eigen-wave equation $\omega^2\mu\bar{\bar{\varepsilon}} \cdot \mathbf{E} + n_d\mathbf{k} \times (n_d\mathbf{k} \times \mathbf{E}) = 0$ resulting from Eq. (1) [13]. Further, we have $\mathbf{D}//Z \Rightarrow \mathbf{D}_x = \mathbf{D}_y = 0$ $\Rightarrow \mathbf{E}_x = \mathbf{E}_y = 0 \Rightarrow \mathbf{D} = \varepsilon_z \mathbf{E}_z$; namely $\mathbf{E}//\mathbf{D}$ also holds for $n_d\mathbf{k} \perp Z$.